\documentclass[a4paper,10pt,fleqn]{article}
\begin{document}
\def\diag{\mathop{\rm diag}}
\def\lam#1#2{\lambda{#1}({#2})}
\def\mi#1#2{\mu{#1}{(#2)}}
\def\p#1#2{\frac{\partial#1}{\partial#2}}
\def\lap{\triangle}
\def\Del{\Delta}
\def\ept{\epsilon(t)+p(t)}
\def\epeta{\epsilon(\eta)+p(\eta)}
\def\epzeta{\epsilon(\zeta)+p(zeta)}
\def\squarebox#1{\mathop{\mkern0.5\thinmuskip
\vbox{\hrule\hbox{\vrule\hskip#1\vrule height#1 width
0pt\vrule}\hrule} \mkern0.5\thinmuskip}} 
\def\blacksquare{\vrule height 1.1ex width 1.1ex depth -.0ex}
\def\dalam{\squarebox{1.4ex}}
\def\etax{(\eta,{\bf x})}
\def\zetax{(\zeta,{\bf x})}
\def\tx{(t,{\bf x})}
\def \bfx{{\bf x}}
\def\veck{\bf k}
\def \kx{{\veck \cdot \bfx}}
\def\element33{~\frac{\sin^2(\sqrt{K}\chi)}{K}}
\def\sound{{c_{\rm s}}}
\def\grant{No.~2~P03D~014~17}
\def \Journal#1#2#3#4{{#1} {\bf #2} (#3) #4}
\def \AP{Adv. Phys.}
\def \ApJ{Astrophys. J.}
\def \JETP{Sov. Phys. JETP}
\def \MNRAS{Mon. Not. R. Astron. Soc.}
\def \ASS{Astroph. Space Sci.}
\def \CQG{Class. Quantum Grav.}
\def \JMP{J. Math. Phys.}
\def \PR{Phys. Rep.}
\def \PRD{{Phys. Rev.} D}
\def \PRL{Phys. Rev. Lett.}
\def\PACS{\par\leavevmode\hbox {\it PACS:\ }}%
\renewcommand{\thefootnote}{\fnsymbol{footnote}}
\setcounter{footnote}{1}

\begin{center}
{\Large\bf A field theory approach to cosmological density 
perturbations\footnote{\sl Article published in Physics Letters A 310 (2003) 357--362}}
\end{center}
\bigskip
\begin{center}
Zdzis{\l}aw A.~Golda and Andrzej Woszczyna\\
\smallskip
{\em Astronomical Observatory, Jagellonian University\\
ul. Orla 171, 30--244 Krak\'ow, Poland}
\end{center}
\medskip

\renewcommand{\thefootnote}{\arabic{footnote}}
\setcounter{footnote}{0}

\begin{abstract}
Adiabatic perturbations propagate in the expanding universe  
like scalar massless fields in some effective Robertson-Walker 
space-time.
\end{abstract}

%\leftline{\small \it Key words: \rm  universe, density perturbation, scalar field, acoustic geometry}
%\leftline{\small \PACS 9880H}

%\addtocounter{page}{0}
\section{Introduction}

Sound propagation in rotationless fluid can be identified
with the propagation of scalar field in some pseudo-Riemannian
space. The geometry of this space is called the acoustic geometry.
In the particular case of steady flows one obtains Unruh
metric, which mimics Schwarzschild
space-time~\cite{Unruh1981&Unruh1995&Visser&Bilic}. 
Potentially, some basic properties of black holes can be
examined in the laboratory by investigating their hydrodynamic
analogues. On the other hand, several theoretical problems with
moving fluids (for instance the acoustic energy problem in the
expanding medium) can find their natural description in
geometrical language~\cite{Stone}.

Similar geometrization can be performed for the density
perturbations in the expanding universe. This fact was
noted for the radiation-dominated and spatially flat Friedman 
universe by Sachs, Wolfe~\cite{Sachs&Wolfe}, Field and
Shepley~\cite{Field&Shepley}. The purpose of this paper is to
point out that the theorem proved by Sachs-Wolfe~(\cite{Sachs&Wolfe}, p. 76 ) can 
be extended to all adiabatic perturbations, and to the universe 
of arbitrary space curvature. We demonstrate this in
synchronous Lifshitz formalism. We show (in the Appendix) that
all major gauge-invariant formalisms admit variables similar to the
Field-Shepley $H$-variable, or the Sachs-Wolfe $E$-variable, i.e. the
quantities, which propagate like the scalar field or gravitational 
waves on some effective Robertson-Walker background.

Throughout this paper Greek indices run from 0 to 3, 
while the convection $c=1$ and $8\pi G=1$ is used.

\section{Synchronous system of reference}

Consider  Friedman-Robertson-Walker universe with the metric form
\begin{equation}
	g_{\mu\nu}=a^2(\eta)\,
	\diag
\left[-1,~1,~\element33,~\element33\sin^2\vartheta\right],
	\label{eq:geom_aku01}
\end{equation}		
and the perfect-fluid energy-momentum tensor 
	\begin{equation}
T^{\mu\nu}=\left[\epsilon +p(\epsilon )\right]u^\mu u^\nu+p(\epsilon )\,g^{\mu\nu}.
	\label{eq:tensor_cieczy}
	\end{equation}	
The equation of state $p=p(\epsilon)$,  assumed here, limits
investigations to the adiabatic density perturbations
(see~\cite{Bardeen} section V{$\!$}.A).  
In particular dissipative processes are excluded.
When the synchronous reference system is used the metric corrections 
caused by the scalar perturbations are determined by two scalar functions
$\lambda(\eta)$ and $\mu(\eta)$. Both $\lambda(\eta)$ and $\mu(\eta)$ solve
the system of two second order differential equations~\cite{Lifshitz&Khalatnikov}
	\begin{eqnarray}
\lam{''}{\eta}+2\frac{a'(\eta)}{a(\eta)}\lam{'}{\eta}-\frac{k^2-K}{3}
	\left[\lam{}{\eta}+\mi{}{\eta}
	\right]
&=&0,
	\label{eq:lam}\\[2ex]
\mi{''}{\eta}+\left[2+3\frac{p'(\eta)}{\epsilon'(\eta)}\right]\frac{a'(\eta)}{a(\eta)}\mi{'}{\eta}
+\frac{k^2-4K}{3}\left[1+3\frac{p'(\eta)}{\epsilon'(\eta)}\right]
	\left[\lam{}{\eta}+\mi{}{\eta}
	\right]
&=&0.
	\label{eq:mu}
	\end{eqnarray}
The density contrast $\delta\etax$ can be found 
in the form of Fourier integral
	\begin{eqnarray}
\delta{\etax}=\int \delta_{k}\, Q(\kx){\rm d}^3k+c.c.
	\end{eqnarray}
with  the Fourier coefficients~\cite{Lifshitz&Khalatnikov}
	\begin{eqnarray}
\delta_k=\frac{1}{3\epsilon(\eta)a^2(\eta)}
	\left[
(k^2-4K)\left[\lam{}{\eta}+\mi{}{\eta}
	\right]
+3\frac{a'(\eta)}{a(\eta)}\mi{'}{\eta}
	\right].
	\end{eqnarray}
$k$ stands for the wave number $k\equiv |{\bf k}|$, while $Q(\kx)$ are scalar harmonics.

We introduce a new perturbation variable $\psi\etax$ by
employing the Darboux transformation of the density contrast 
	\begin{eqnarray}
\psi\etax =a(\eta) H^2(\eta)
 \frac{\partial}{\partial\eta}\frac{\delta{\etax}}{H(\eta) 
\left[1+\frac{p(\eta)}{\epsilon(\eta)}\right]}.\hspace{6.5cm}
	\end{eqnarray}
With system (\ref{eq:lam}--\ref{eq:mu}) satisfied, the propagation equation
for the variable $\psi\etax $ takes the form 
	\begin{eqnarray}
\p{^2}{\eta^2}\psi{\etax}+ 
	\left(
2\frac{{\sf a}'(\eta)}{{\sf a}(\eta)}-\frac{\sound'(\eta)}{\sound(\eta)}
	\right)
\p{}{\eta}\psi{\etax}-\sound^2(\eta)\, \lap\psi{\etax}=0,
	\label{eq:fala2}
	\end{eqnarray}
where $\lap$ is Laplacian operating in the constant time hypersurface, 
	\begin{eqnarray}
\sound(\eta)=\sqrt{\frac{p'(\eta)}{\epsilon'(\eta)}}\quad\mbox{and}\quad 
{\sf a} (\eta)=a(\eta)\sqrt{\frac{1}{\sound(\eta)} 
\frac{\epsilon(\eta)+p(\eta)}{3H^2(\eta)}}.
	\end{eqnarray}
$\sound(\eta)$ stands for the sound velocity, which is assumed to be
strongly positive. $H(\eta)=a'(\eta){\slash}a^2(\eta)$ stands for the
Hubble parameter. Although the gauge modes may contribute to the
density contrast $\delta{\etax}$, the new variable $\psi$ is
gauge invariant\footnote{The construction of the gauge invariant
variables by means of Darboux transformations in a radiation
dominated universe has been discussed in detail
in~\cite{Golda&Woszczyna}, and independently employed
in~\cite{Grishchuk}.}. The equation~(\ref{eq:fala2}) has the
same structure as the Lukash~\cite{Lukash1980} equation~(1.10)
and can be independently derived on the ground of the Lagrangian
formalism.

Let us introduce a new time variable $\zeta$ defined by the integral
	\begin{eqnarray}
\zeta =\int\!\! \sound(\eta)\,{\rm d} \eta.
	\label{eq:dzeta}
	\end{eqnarray}
The change of the time variable allows one to
reduce the propagation equation for $\psi$ to
	\begin{eqnarray}
\p{^2}{\zeta^2}\psi{\zetax} +2\frac{{\sf a}'(\zeta)}{{\sf a}(\zeta)}
\p{}{\zeta}\psi {\zetax}- \lap \psi{\zetax}=0,
	\label{eq:fala3}
	\end{eqnarray}
which is an explicit form of the d'Alambert equation
	\begin{eqnarray}
\dalam\psi{\zetax} \equiv \nabla^\mu\nabla_\mu\psi{\zetax}=0
	\label{eq:Klein-Gordon}
	\end{eqnarray}
for scalar field $\psi$ propagating in 
Robertson-Walker space-time  
	\begin{equation}
{\sf g}_{\mu\nu}={\sf a}^2(\zeta)\,
\diag 
	\left[-1,~1,~\element33,~\element33\sin^2\vartheta\right]
	\label{eq:geom_aku02}
	\end{equation}		
with a scale factor ${\sf a}(\zeta)$ 
	\begin{equation}
{\sf a}(\zeta)=a(\zeta)
\sqrt{\frac{1}{\sound(\zeta)}\frac{\epsilon(\zeta)+p(\zeta)}{3H^2(\zeta)}}.
	\label{eq:nowy_czynnik}
	\end{equation}  
The equation (\ref{eq:Klein-Gordon}) with the metric tensor
(\ref{eq:geom_aku02}) build the acoustic geometry, we looked
for. In consequence, the adiabatic perturbations propagate as
massless scalar field in the expanding universe.  The new time
variable $\zeta$ plays a similar role in the acoustic geometry
as does the conformal time $\eta$  in the original universe.
This particularly concerns the shape of null cones and the sound
horizons, therefore, $\zeta$  may be considered as the {\it
acoustic conformal time}.  Although the d'Alambert operator is
Lorentz-invariant the propagation
equation~(\ref{eq:Klein-Gordon}) is not, because of
non-invariance of the field variable $\psi$. This limitation is
characteristic of quasi-particles (phonons --- in cosmological
context see~\cite{Chibisov&Mukhanov}) and well-known in the
acoustic geometrical
descriptions~\cite{Unruh1981&Unruh1995&Visser&Bilic}.

The reduction of the density perturbation equation to the
d'Alambert equation, presented above, may also be understood as
a generalized Sachs-Wolfe theorem~(\cite{Sachs&Wolfe}, p 76 ).  Equation
(\ref{eq:Klein-Gordon}) (and in consequence (\ref{eq:fala3})) is
identical (up to some constant tensor factor) with the equation
for gravitational wave propagating in the pressureless
environment~\cite{Grishchuk}.  Similar equations for the density
perturbations can be obtained in all major gauge invariant
formalisms. Appropriate Darboux transformations for
gauge-invariant equations are given in Appendix.

\section{Summary}

For adiabatic perturbations we 
constructed the perturbation variables ($\psi$ in the synchronous system, and its 
analogues in gauge-invariant formalisms), which propagate like massless scalar fields 
(\ref{eq:Klein-Gordon}) in some Robertson-Walker space-time. The construction of 
these fields and the background geometry is unique and similar in the most frequently used 
gauge-invariant descriptions. Since the propagation equation has been reduced to d'Alambert 
equation, the field theory in the curved space-time~\cite{Birrell&Davies} become an suitable 
language to describe density perturbations and to determine their visual effects on the cosmic 
microwave background.

\section*{Acknowledgements}
This paper was inspired by lectures on the black-hole acoustic geometry given by M. Visser 
on ERE-2001 (Universidad Complutense, Madrid 2001) and by  W. Unruh on COSLAB School 
(Jagellonian University, Krak\'ow 2002). The authors thank Marek Demia\'nski for fruitful discussions and 
critical remarks. This work was partially supported by State Committee for
Scientific Research, project  \grant.

\section*{Appendix}

In this section we list transformations of gauge-invariant
propagation equations to the form of equation~(\ref{eq:fala2}).
We limit ourselves to the transformations of perturbation
variables --- the time transformation in all cases is identical
and equal to~(\ref{eq:dzeta}).  We hold the original authors'
notation where possible, yet to reach the consistency of this
paper we always denote the metric scale factor by $a(t)$, the
energy density by $\epsilon$, the sound velocity by $\sound$,
the expansion rate by Hubble
$H=a'(t){\slash}a(t)=a'(\eta){\slash}a^2(\eta)$ or the expansion
scalar $\theta=3H$. Capital $K$ stands for the curvature index
$K=0,-1,1$, while small $k$ denotes the wave number.

It should be emphasized that although the quantities
$\widehat{\Del\epsilon}$, $\widehat\Del$, $\widehat{\Phi_{\rm
H}}$, $\widehat{\epsilon_{\rm m}}$, $\widehat{\delta\epsilon}$,
$\widehat\phi$, below, obey the same propagation equation, they
should not be identified. Geometrical differences between them
are clear from how the corresponding ${\Del\epsilon}$, $\Del$,
${\Phi_{\rm H}}$, ${\epsilon_{\rm m}}$, ${\delta\epsilon}$,
$\phi$ are defined in the original papers.

\bigskip
\noindent
{\em Author\/}: Olson~\cite{Olson} (generalized to arbitrary $K$ in \cite{Woszczyna&Kulak}) \\
{\em Propagation equation\/} (\cite{Woszczyna&Kulak} system (21a--21b)):
	\begin{eqnarray}
\p{}{\eta}\Del\epsilon{\etax}&=&a(\eta)
	\left[
-\frac53\theta(\eta)\Del\epsilon{\etax}-\left[\epsilon(\eta)+p(\eta)\right]
\Del\theta{\etax}
	\right],\\
\p{}{\eta}\Del\theta{\etax}&=&a(\eta)
	\left[
-\frac{\sound^2(\eta)}{a^2(\eta)\left[\epsilon(\eta)+p(\eta)\right]}
\left[\lap+3K\right]\Del\epsilon{\etax}-
\frac12\Del\epsilon{\etax}\right.\nonumber\\
&&\left.{}-\frac43\theta(\eta)\Del\theta{\etax}
	\right].
	\end{eqnarray}
{\em Variables\/}:\\
$\Del\epsilon$ --- the spatial density Laplacian, \\
$\Del\theta$ --- the spatial Laplacian of the expansion rate.\\ 
The wave equation $\dalam \widehat{\Del\epsilon}{\etax}=0$,
given explicitly by eq. (\ref{eq:fala2}), (equivalently the 
equation (\ref{eq:Klein-Gordon}) expressed in the {\it acoustic conformal time}) 
is satisfied by:
\begin{eqnarray}
\widehat{\Del\epsilon}{\etax}=\frac{H^2(\eta)}{a^2(\eta)
	\left[
	\epeta
	\right]
}\,\p{}{\eta}
	\left[
\frac{a^5(\eta)}{H(\eta)}\Del\epsilon
{\etax}
	\right].
	\end{eqnarray}

\bigskip
\noindent
{\em Authors\/}: Bruni, Dunsby and Ellis (\cite{Bruni&Dunsby&Ellis} and references given there)\\
{\em Propagation equation\/}  (\cite{Bruni&Dunsby&Ellis} equation~(73)):
	\begin{eqnarray}
\p{^2}{t^2}\Del\tx+(2+3\sound^2-6w)H\p{}{t}{\Del}\tx
-	\left[
	\left(
	\frac12+4w-\frac32w^2-3\sound^2
	\right)\kappa\,\epsilon\right.\nonumber\\
\left.+ 12(\sound^2-w)\frac{K}{a^2} 
	\right]\Del\tx
-\sound^2\,{}^{(3)}\nabla^2\Del\tx =0.
	\label{eq:ellis&bruni&hwang&dunsby}
	\end{eqnarray}
{\em Variables\/}:\\
$w=p\slash\epsilon$ --- pressure to energy ratio,\\
perturbation variable capital delta: $\Del{\etax}=a^2(\eta) \frac{\Del\epsilon{\etax}} {\epsilon(\eta)}$.\\
The wave equation $\dalam\widehat{\Del}{\etax}=0$ satisfied by:
	\begin{eqnarray}
\widehat{\Del}{\etax}=\frac{H^2(\eta)}{a^2(\eta)
	\left[
\epeta
	\right]}\,
\p{}{\eta}
	\left[
\frac{a^3(\eta)\epsilon(\eta)}{H(\eta)}\Del{\etax}
	\right].
	\end{eqnarray}

\bigskip
\noindent
{\em Author\/}: Bardeen~\cite{Bardeen}\\
{\em Propagation equation\/} (not given explicitly in the cited paper):
	\begin{eqnarray}
\p{^2}{\eta^2}\Phi_{\rm H}{\etax} +\left[1+\sound^2(\eta)\right]
a(\eta)\theta(\eta)\p{}{\eta}\Phi_{\rm H}{\etax}\nonumber\\ 
-
	\left[
2 K\left[1+3\sound^2(\eta)\right]+a^2(\eta)\epsilon(\eta)
	\left[
\frac{p(\eta)}{\epsilon(\eta)}-\sound^2(\eta)
	\right]
+\sound^2(\eta)\lap 
	\right]
\Phi_{\rm H}{\etax}=0.
	\end{eqnarray}
{\em Variables\/}:\\ $\Phi_{\rm H}$ --- gauge invariant potential.\\
 The wave equation $\dalam \widehat{\Phi_{\rm H}}{\etax}=0 $ satisfied by 
	\begin{eqnarray}
\widehat{\Phi_{\rm H}}{\etax} =\frac{H^2(\eta)}{a^2(\eta)
	\left[
\epeta
	\right]}\,
\p{}{\eta}
	\left[
\frac{a(\eta)}{H(\eta)}\Phi_{\rm H}{\etax}
	\right].
	\end{eqnarray}

\bigskip
\noindent
{\em Author\/}: Bardeen~\cite{Bardeen}\\
{\em Propagation equation\/}~(\cite{Bardeen}~equation~(4.9))
	\begin{eqnarray}
&&\p{^2}{\eta^2}
	\left[
\epsilon(\eta) a^3(\eta)\epsilon_{\rm m}{\etax}
	\right]
+
	\left[
1+ 3\sound^2(\eta )
	\right]
\frac{a'(\eta)}{a(\eta)} \p{}{\eta} 	
	\left[
\epsilon(\eta) a^3(\eta)\epsilon_{\rm m}{\etax}
	\right]\nonumber\\
&&+
	\left[
-\sound^2(\eta)\left[\lap+3K\right]-\frac12
	\left[
\epsilon(\eta)+p(\eta)
	\right]a^2(\eta)
	\right]	
	\left[
\epsilon(\eta) a^3(\eta)\epsilon_{\rm m}{\etax}
	\right]
=0.
	\label{bardeen}
	\end{eqnarray}
{\em Variables\/}:\\ $\epsilon_{\rm m}$ --- the density contrast
measured on the flow-orthogonal surfaces.\\ 
The wave equation $\dalam \widehat{\epsilon_{\rm m}}{\etax}=0$ satisfied by:
	\begin{eqnarray}
\widehat{\epsilon_{\rm m}}{\etax}=\frac{H^2(\eta)}{a^2(\eta)
	\left[
\epeta
	\right]}\,
\p{}{\eta}
	\left[
\frac{a^3(\eta)\epsilon(\eta)}{H(\eta)}{\epsilon}_{\rm m}{\etax}
	\right].
	\end{eqnarray}

\bigskip
\noindent
{\em Authors\/}: Lyth and Stewart~(\cite{Lyth&Stewart} and
references given there)\\ 
{\em Propagation equation\/}~(\cite{Lyth&Stewart},~equations~(35--36))
	\begin{eqnarray}
\p{}{t}\delta\epsilon\tx&=&
-\theta(t)\delta\epsilon\tx-
	\left[
\ept
	\right]
\delta\theta\tx,\\ 
\p{}{t}\delta\theta\tx&=&-
	\left[
\frac12+\frac{\sound^2(\eta) 
	\left[
3\epsilon(t)-\theta^2(t)
	\right]
}{3	
	\left[\ept
	\right]
}
	\right]
\delta\epsilon\tx-\frac{2}{3}\theta(t)\delta\theta\tx\nonumber\\
&&-\frac{\sound^2(\eta)}{\ept}\lap\delta\epsilon\tx.
	\label{lyth}
	\end{eqnarray}
{\em Variables\/}:\\ 
$\delta\epsilon$, $\delta\theta$ --- variations of density 
and the expansion rate on the flow-orthogonal hypersurfaces.\\
The wave equation $\dalam \widehat{\delta\epsilon}{\etax}=0$ satisfied by:
	\begin{eqnarray}
\widehat{\delta\epsilon}{\etax}=\frac{H^2(\eta)}{a^2(\eta)
	\left[
\epeta
	\right]}\,
\p{}{\eta}
	\left[
\frac{a^3(\eta)\epsilon(\eta)}{H(\eta)}\delta\epsilon{\etax}
	\right].
	\label{eq:transf_{Lyth&Stewart}}
	\end{eqnarray}

\bigskip
\noindent
{\em Authors\/}: Brandenberger, Kahn and Press~\cite{Brandenberger&Kahn&Press}
(generalized to arbitrary $K$ in \cite{Mukhanov&Feldman&Brandenberger})\\
{\em Propagation equation\/}  (longitudinal gauge, \cite{Mukhanov&Feldman&Brandenberger} equation~(5.3) for $\delta S=0$.) 
	\begin{eqnarray}
\phi''{\etax}&+&3(1+\sound^2(\eta)){\mathcal H}(\eta)\phi'{\etax}
-\sound^2(\eta)\lap\phi{\etax}\nonumber\\
&+&	\left[
2{\mathcal H}'(\eta)+(1+3 \sound^2(\eta)) 
({\mathcal H}^2(\eta)-K)
	\right]
\phi{\etax}=0.
	\end{eqnarray}
{\em Variables\/}:\\
${\mathcal H}(\eta)=a'(\eta)\slash a(\eta)$,\\
$\phi{\etax}$ --- gauge-invariant potential.\\
The wave equation $\dalam \widehat{\phi}{\etax}=0$ satisfied by:
	\begin{eqnarray}
\widehat{\phi}{\etax}=\frac{1}{\epeta}
	\left[
\frac{{\mathcal H}(\eta)}{a^2(\eta)} 
	\right]^2
\p{}{\eta}
	\left[
\frac{a^2(\eta)}{{\mathcal H}(\eta)}\phi{\etax}
	\right].
	\label{eq:transf_{Brandenberger}}
	\end{eqnarray}

\bibliographystyle{plain}

\end{document}